\documentclass[aps,a4paper,twocolumn,nofootinbib]{revtex4}
\usepackage{amsmath,amssymb,graphicx,color,dsfont}

\usepackage{pdfpages}

\newcommand{\integ}[1]{\ensuremath{\int \!\! \mathrm{d}#1 \,}}

\newcommand{\bra}[1]{\ensuremath{\left\langle #1 \right|}}
\newcommand{\ket}[1]{\ensuremath{\left| #1 \right\rangle}}

\newcommand{\PL}{\ensuremath{P_L}}
\newcommand{\XM}{\ensuremath{X_M}}
\newcommand{\PM}{\ensuremath{P_M}}

\newcommand{\PLOne}{\ensuremath{P_L^{(p_1)}}}
\newcommand{\PLTwo}{\ensuremath{P_L^{(p_2)}}}
\newcommand{\PLRead}{\ensuremath{P_L^{(r)}}}

\newcommand{\PLout}{\ensuremath{P_L^{\textrm{out}}}}
\newcommand{\PLin}{\ensuremath{P_L^{\textrm{in}}}}
\newcommand{\XLout}{\ensuremath{X_L^{\textrm{out}}}}
\newcommand{\XLin}{\ensuremath{X_L^{\textrm{in}}}}

\newcommand{\PMout}{\ensuremath{P_M^{\textrm{out}}}}
\newcommand{\PMin}{\ensuremath{P_M^{\textrm{in}}}}
\newcommand{\XMout}{\ensuremath{X_M^{\textrm{out}}}}
\newcommand{\XMin}{\ensuremath{X_M^{\textrm{in}}}}

\newcommand{\mean}[1]{\ensuremath{\langle #1 \rangle}}

\newcommand{\prob}[1]{\textrm{Pr} \! \left(#1\right)}

\newcommand{\nbar}{\ensuremath{\bar{n}}}
\newcommand{\nbareff}{\ensuremath{\bar{n}_{\textrm{eff}}}}

\newcommand{\xzp}{\ensuremath{x_0}}

\newcommand{\omegam}{\ensuremath{\omega_M}}

\newcommand{\meff}{\ensuremath{m_{\textrm{eff}}}}
\newcommand{\Teff}{\ensuremath{T_{\textrm{eff}}}}

\newcommand{\etal}{\emph{et al}.}

\newcommand{\CQG}[3]{Class. Quantum Grav.~\textbf{#1}, #2 (#3)}

\newcommand{\EPJD}[3]{Eur. Phys. J. D~\textbf{#1}, #2 (#3)}
\newcommand{\JETPLett}[3]{JETP Lett. \textbf{#1}, #2 (#3)}

\newcommand{\Nature}[3]{Nature (London)~\textbf{#1}, #2 (#3)}
\newcommand{\NatPhys}[3]{Nature Physics~\textbf{#1}, #2 (#3)}
\newcommand{\NatPhot}[3]{Nature Photonics~\textbf{#1}, #2 (#3)}
\newcommand{\NatComm}[3]{Nat. Commun.~\textbf{#1}, #2 (#3)}
\newcommand{\NJP}[3]{New J. Phys.~\textbf{#1}, #2 (#3)}
\newcommand{\Physics}[3]{Physics~\textbf{#1}, #2 (#3)}
\newcommand{\PNAS}[3]{Proc. Natl. Acad. Sci. USA \textbf{#1}, #2 (#3)}
\newcommand{\PRA}[3]{Phys. Rev. A~\textbf{#1}, #2 (#3)}
\newcommand{\PRAR}[3]{Phys. Rev. A~\textbf{#1}, #2(R) (#3)} 

\newcommand{\PRL}[3]{Phys. Rev. Lett.~\textbf{#1}, #2 (#3)}
\newcommand{\PRX}[3]{Phys. Rev. X~\textbf{#1}, #2 (#3)}
\newcommand{\RMP}[3]{Rev. Mod. Phys.~\textbf{#1}, #2 (#3)}
\newcommand{\Science}[3]{Science~\textbf{#1}, #2 (#3)}

\begin{document}

\title{Cooling-by-measurement and mechanical state tomography via pulsed optomechanics}



\author{M.~R.~Vanner}\email[E-mail: ]{michael.vanner@univie.ac.at}
\author{J.~Hofer}
\author{G.~D.~Cole}
\author{M.~Aspelmeyer}\email[E-mail: ]{markus.aspelmeyer@univie.ac.at}
\affiliation{
\mbox{Vienna Center for Quantum Science and Technology (VCQ),}\\ 
\mbox{Faculty of Physics, University of Vienna, Boltzmanngasse 5, A-1090 Vienna, Austria}}

\date{First submitted: November 29, 2012; Latest revision: \today}


\begin{abstract}
Observing a physical quantity without disturbing it is a key capability for the control of individual quantum systems. Such back-action-evading or quantum-non-demolition measurements were first introduced in the 1970s in the context of gravitational wave detection to measure weak forces on test masses by high precision monitoring of their motion. Now, such techniques have become an indispensable tool in quantum science for preparing, manipulating, and detecting quantum states of light, atoms, and other quantum systems. Here we experimentally perform rapid optical quantum-noise-limited measurements of the position of a mechanical oscillator by using pulses of light with a duration much shorter than a period of mechanical motion. Using this back-action evading interaction we performed both state preparation and full state tomography of the mechanical motional state. We have reconstructed mechanical states with a position uncertainty reduced to 19~pm, limited by the quantum fluctuations of the optical pulse, and we have performed `cooling-by-measurement' to reduce the mechanical mode temperature from an initial 1100~K to 16~K. Future improvements to this technique may allow for quantum squeezing of mechanical motion, even from room temperature, and reconstruction of non-classical states exhibiting negative regions in their phase-space quasi-probability distribution.
\end{abstract}

\maketitle



Experiments are now beginning to investigate non-classical motion of massive mechanical devices~\cite{OConnell2010, Lee2011Entangle, Lee2012NonClass}. This opens up new perspectives for quantum-physics enhanced applications and for tests of the foundations of physics. A versatile approach to manipulate mechanical states of motion is provided by the interaction with electromagnetic radiation, typically confined to microwave or optical cavities. Such cavity-optomechanics experiments~\cite{Blencowe2004, Schwab2005, Kippenberg2008, Marquardt2009, Aspelmeyer2012} have thus far largely concentrated on high sensitivity continuous monitoring of the mechanical position~\cite{Briant2003, Regal2008, Abbott2009, Teufel2011, Chan2011, Faust2012}. Because of the back-action imparted by the probe onto the measured object, the precision of such a measurement is fundamentally constrained by the standard quantum limit (SQL)~\cite{BraginskyBook, LaHaye2004}, and therefore only allows for classical phase-space reconstruction~\cite{Rugar1991, Tittonen1999, Briant2003}. In order to observe quantum mechanical features that are smaller than the mechanical zero-point motion, back-action-evading measurement techniques that can surpass the SQL~\cite{ref:QND,Braginsky1980} are required. Following the early insights of Braginsky, beating the standard quantum limit \emph{`can be achieved only in one way: design the probe so it ``sees'' only the measured observable'}~\cite{BraginskyBook}.
Such back-action evading techniques were first realized for the detection of optical quadratures~\cite{Levenson1986, LaPorta1989, Grangier1998} and have since been used for, e.g. precision measurement of atomic ensemble spin \cite{Kuzmich2000, Teper2008, Takano2009, Appel2009, SchleierSmith2010} and quantum non-demolition microwave photon counting~\cite{Guerlin2007}. In optomechanics, to perform a back-action evading measurement of the mechanical position, a time-dependent measurement scheme is required. One prominent example is the so-called `two-tone approach'~\cite{Braginsky1980, Clerk2008}, which uses a probe with an intensity that oscillates at twice the mechanical frequency. The field probes the mechanics periodically and the back-action imparted to the mechanical motion by the optical probe does not affect the measurement of the mechanical amplitude. This is closely analogous to a stroboscopic measurement of the mechanical motion~\cite{Braginsky1980}. Using the two-tone approach with a microwave probe field, a back-action evading interaction was recently realized to measure a single quadrature of a nanomechanical resonator~\cite{Hertzberg2010}.

Here we take a different tack to perform position measurements of a mechanical oscillator using single optical pulses. Our experimental approach employs optical pulses that have a duration much shorter than a mechanical period of motion. This provides a back-action-evading interaction for measuring the mechanical position because the interaction leaves the position unchanged, perturbing only the mechanical momentum, and was first suggested by Braginsky~\cite{Braginsky1978}. The precision of this pulsed measurement process is no longer limited by the SQL but is ultimately limited by the quantum optical phase noise. We implement the pulsed protocol proposed in Ref.~\cite{VannerPNAS2011} where one or two pulses are used to prepare a motional state `by measurement' and then a subsequent pulse is used for state tomography. Mechanical state preparation `by measurement' is achieved by utilising the information gained from the pulsed measurement to update the probability distribution that describes the motional state. The experiments reported here have been performed in the weak interaction regime where the back action itself is negligible, however, the pulsed measurements have a dramatic effect to the mechanical thermal state and the measurement precision we achieved was set by the quantum optical phase noise. We therefore require a quantized description of the optical field, however, at this stage, the mechanical motional state may be described classically. Our protocol can be used to prepare mechanical states independent of the initial mechanical thermal occupation and thus, no pre-cooling of the mechanical motion is required. Moreover, our pulsed protocol has the advantage that the experiment can be performed on a timescale faster than decoherence or rethermalization and thus has considerable tolerance to the surrounding thermal bath~\cite{VannerPNAS2011}. Employing our pulsed approach, mechanical dynamics rather than the steady-state can be conveniently probed and non-equilibrium mechanical behavior can be characterized. Also note that pulsed quantum optomechanics operates fully in the so-called `non-resolved sideband regime', in which the cavity decay rate is much larger than the mechanical frequency. Indeed, all results reported here were obtained without the use of an optical cavity.


\section*{Experimental protocol}

\begin{figure}[!ht]
\includegraphics[width=1.0\hsize]{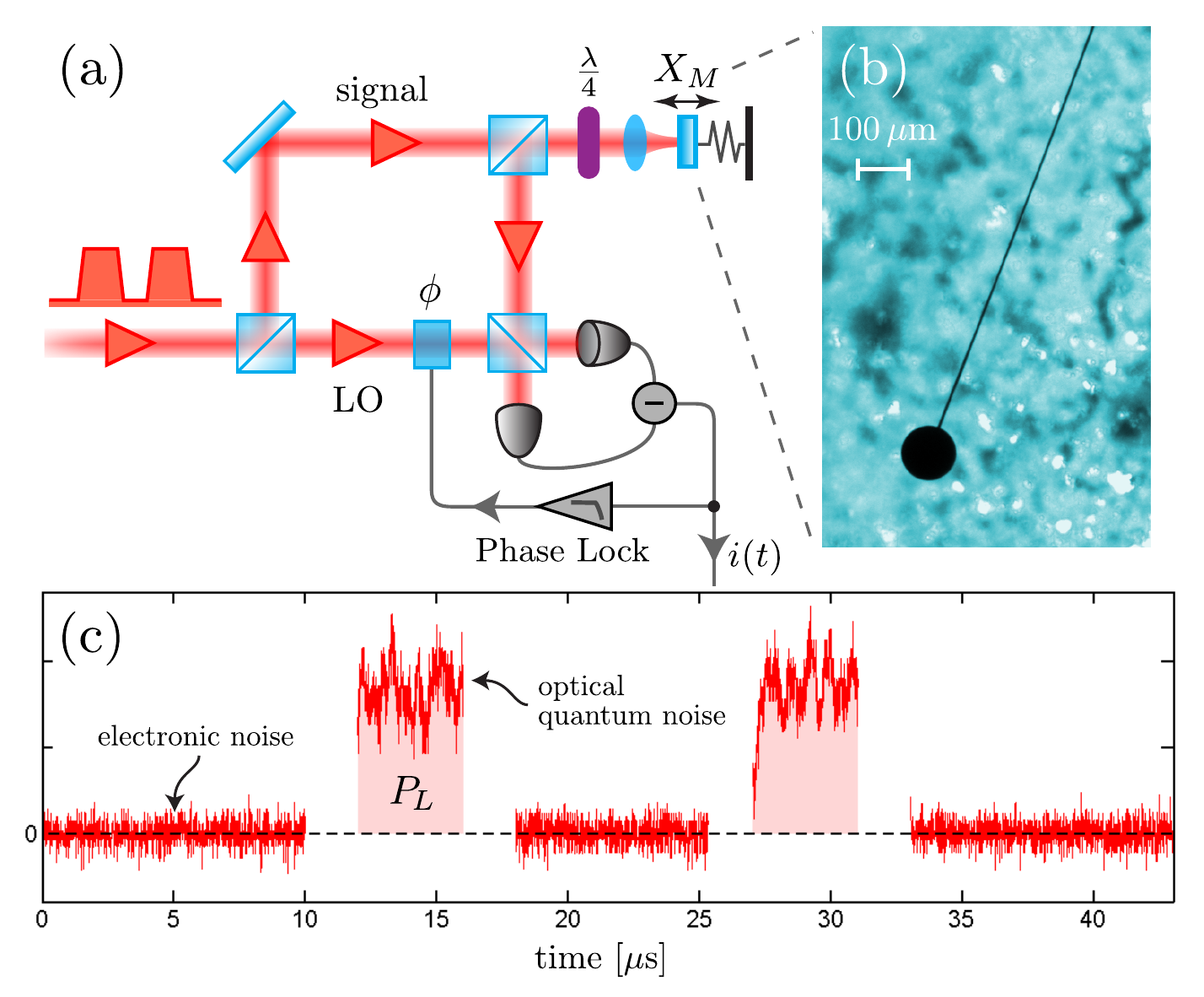}
\caption{
(a) Schematic of the experimental setup used to perform state tomography and state preparation of the motional state of a mechanical resonator. In addition to the optical pulses a weak continuous field is used to stabilize the interferometer phase using the homodyne output passed through a low-pass filter with cutoff frequency below the mechanical frequency. (b) Colourized optical micrograph of the high-reflectivity micro-mechanical cantilever fabricated for this experiment. The head of the cantilever, where the signal beam is focussed, is 100~$\mu$m in diameter. (c) Example time trace of the homodyne output for a pair of 4~$\mu$s pulses. (For clarity, the pulse rising and falling edges are not shown.) The measurement outcome $\PL$ is the time integral of the homodyne output (indicated by the shaded region). Time resolved optical quantum noise is visible during the pulse.}
\label{Fig:Setup}
\end{figure}

Our experimental setup is shown schematically in Fig.~\ref{Fig:Setup}(a). Optical pulses are injected into a Mach-Zehnder interferometer that has a micro-mechanical oscillating mirror in one of the two interferometer paths. The pulses are first divided by a beam-splitter that forms one intense beam that acts as a local oscillator (LO) and one weak beam that we will henceforth refer to as the signal. The signal is focussed onto and reflects from a micro-mechanical oscillator (Fig.~\ref{Fig:Setup}(b)). During the reflection of the short optical pulse, changes to the position of the mechanical oscillator are negligible. The coherent optical pulse gains a phase shift in proportion to the mechanical position, which is accurately described by a phase quadrature displacement as the  mechanical position fluctuations are small. Concurrently, the radiation-pressure force of the reflection imparts momentum to the mechanical resonator. This momentum can be decomposed into a classical component due to the mean photon number and a component dependent upon the photon number fluctuations. Quantitatively, this optomechanical interaction is described by the input-output relations
\begin{align}
\XLout = \XLin, \qquad & \PLout =  \PLin + \chi \XMin, \notag\\
\XMout = \XMin, \qquad & \PMout =  \PMin + \chi \XLin + \Omega.
\end{align}
Here, the subscripts label the light ($L$) and mechanics ($M$); $X$ and $P$ are the dimensionless amplitude (position) and phase (momentum) quadratures for the light (mechanics); $\chi = 4 \pi \xzp \sqrt{N}/\lambda$ quantifies the quadrature information exchanged between the light and the mechanics and determines the strength of the mechanical position measurement, and $\Omega = 8\pi \xzp N/\lambda$ is the classical momentum transfer to the mechanical oscillator. ($N$, mean photon number per pulse; $\lambda$, optical wavelength; $\xzp = (\hbar/2 \meff \omega_M)^{1/2}$, mechanical ground state extension; $\meff$, mechanical effective mass; $\omega_M$, mechanical angular frequency.) After the optomechanical reflection, the signal then overlaps and interferes with the LO pulse on a 50/50 beam-splitter where the (mean) phase between the LO and signal beams is set to be $\pi/2$. The intensities of both beam-splitter outputs are measured by photodiodes and the photocurrents are subtracted to implement homodyne detection of the optical phase quadrature. A typical difference current time trace is shown in Fig.~\ref{Fig:Setup}(c) where the measurement outcome $\PL$ is the time integral over the pulse duration of the difference current.

After the pulsed measurement the mechanical state of motion is changed as our knowledge of the mechanical position has increased. For an initial thermal state of the mechanical resonator with a large thermal occupation, i.e. $\chi^2(1 + 2\nbar) > 1$, the means and variances of the mechanical quadratures, upon obtaining the measurement outcome $\PL$ are \cite{VannerPNAS2011}:
\begin{align}
\label{eq:mechInOut}
& \mean{\XMout} \simeq \PL/\chi, \qquad \mean{\PMout} =  \Omega, \notag\\
& \,\sigma_{\XMout}^2 \simeq 1/(2\chi^2), \qquad \,\sigma_{\PMout}^2 = (\chi^2 + 1 + 2\nbar)/2,
\end{align}
where $\nbar \simeq k_B T / \hbar \omegam$ is the mean occupation of the mechanical mode when in thermal equilibrium with the environment at temperature $T > \hbar \omegam / k_B$. Notably the information gained from the measurement reduces the mechanical position variance from $\nbar$ to $1/(2\chi^2)$, which does not depend on the initial occupation.
%
%
The resultant state of mechanical motion, following such a measurement, is no longer in thermal equilibrium with the surrounding environment and has a reduced effective thermal occupation $\nbareff = (\sigma_{\XM}^2 \sigma_{\PM}^2)^{1/2}-1/2 \simeq (\nbar/(2\chi^2))^{1/2}$. Moreover, a subsequent pulse performed after one quarter of a period of mechanical harmonic evolution can measure the mechanical momentum at the time of the first pulse to further reduce the effective occupation. This `cooling-by-measurement' method for entropy reduction, i.e. obtaining mechanical position and then momentum information on the initial state, is rapid and has considerable tolerance to both the initial thermal occupation and the surrounding thermal bath~\cite{VannerPNAS2011}. With future experimental improvements, this scheme allows for the generation of high purity and quantum squeezed states of mechanical motion `by measurement'. Due to the resilience to mechanical thermal noise, this scheme may provide a more feasible route to quantum squeezing than achieving a quantum squeezed state via parametric modulation~\cite{Rugar1991, Briant2003}, which can be combined with continuous measurement and feedback~\cite{Szorkovsky2011}.

In our experiment one or two pulses are used to prepare a mechanical state at a known time. Then a read-out pulse is made after time $\theta/\omegam$ of mechanical harmonic evolution to sample the mechanical probability distribution of the $\theta$-rotated quadrature, i.e. a marginal. Repeating this process many times and obtaining the marginals for a large number of mechanical phase-space angles $\theta$ is sufficient to uniquely determine the mechanical quantum state of motion~\cite{Vogel1989}. Quantum state tomography by measurement of the marginals was first realized with optical fields using homodyne interferometry~\cite{Smithey1993} and has now become an indispensable tool in the field of quantum optics~\cite{Lvovsky2009} being applied to other physical systems such as molecular vibration~\cite{Dunn1995}, spin ensembles~\cite{Fernholz2008}, and microwave fields~\cite{Mallet2011}. Here we implement such mechanical state tomography by utilizing the pulsed measurement outcome probability distribution $\prob{\PL} = \integ{\XM} \pi^{-1/2} \exp[-(\PL - \chi \XM)^2] \,\prob{\XM, \theta}$ that contains the mechanical marginals $\prob{\XM, \theta} = \bra{\XM} \rho_M^{\textrm{in}}(\theta) \ket{\XM}$, where $\rho_M^{\textrm{in}}(\theta)$ is the mechanical input state to be reconstructed after time $\theta/\omegam$ of harmonic evolution~\cite{VannerPNAS2011}. In this experiment we prepare and reconstruct mechanical motional states with features that are not smaller than $\chi^{-1}$ and hence, unless otherwise noted, we use the optical measurement outcome distribution as an approximation for the mechanical distribution using the scaled outcome $\PL/\chi$.

\begin{widetext}

\begin{figure}[!h]
\includegraphics[width=1.0\hsize]{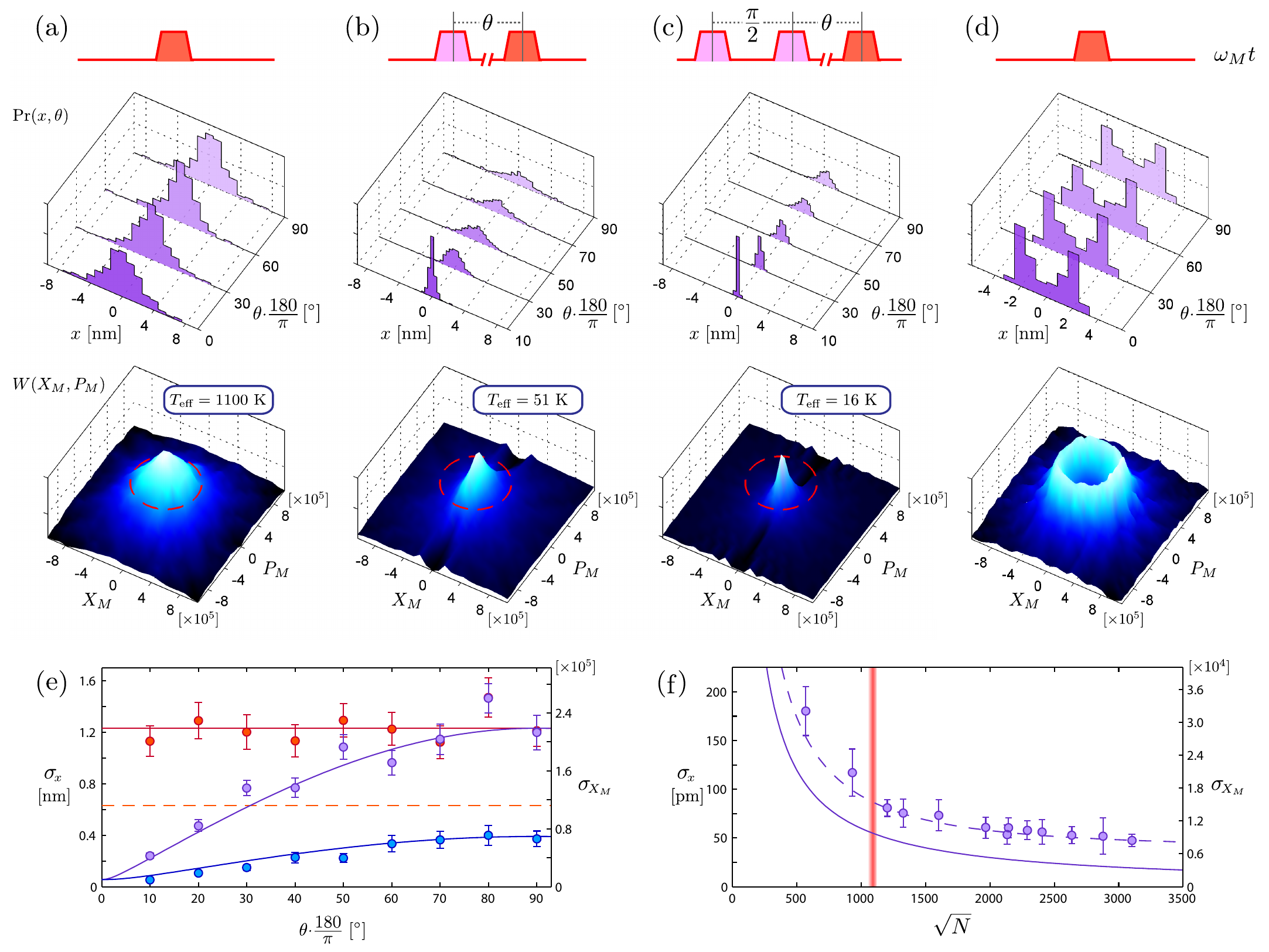}
\caption{
Motional states of a mechanical resonator experimentally prepared and fully reconstructed using optical pulsed quantum measurement. The uppermost row shows the pulsed protocols used; pulses filled pink are used for measurement based state preparation and pulses filled red are used for mechanical state tomography. The two rows below show a subset of the measured probability distributions of the mechanical quadratures $\prob{x,\theta}$, i.e. the marginals, and the reconstructed phase-space distributions $W(\XM, \PM)$, respectively. The phase-space distributions were reconstructed using 9 marginal angles up to $\theta \cdot \pi/180 = 90^{\circ}$ (with a larger number of bins used than that shown for the marginals). To help provide a sense of scale, the marginal distributions are plotted with mechanical displacement in meters and the phase-space distributions are plotted in units of the mechanical zero-point motion.
(a) In the first column tomography and reconstruction of an initial mechanical thermal state driven by white noise up to a mode temperature of 1100~K is shown. The dashed circle has a radius equal to $2\sigma$ of the thermal distribution.
(b) A single pulsed measurement reduces the mechanical position variance, but leaves the momentum distribution unchanged.
(c) `Cooling by measurement' performed with two pulses separated by one quarter of a mechanical period rapidly reduces the mechanical state's entropy. The effective temperature of the mechanical state reconstructed here has been reduced to 16~K.
(d) State reconstruction of a non-Gaussian mechanical state of motion generated by resonant sinusoidal drive.
(e) The (one standard deviation) width of the position distribution observed for states (a-c) with phase-space angle $\theta$. (The left vertical axis is in units of nanometers and the right axis uses units of the mechanical zero-point motion.) The thermal state (red points) shows a position width approximately twice that when at room temperature (dashed line). State (b) has a reduced position width for small phase-space angles (purple points). The position width of state (c) is reduced for all phase-space angles (blue points). The solid lines are theoretical fits obtained using Eq.~(\ref{eq:mechInOut}) generalized for all $\theta$ as well the two-pulse-preparation case.
(f) Plot of the conditional mechanical width with pulse strength measured using preparation and read-out pulses separated by $5^\circ$ of mechanical evolution. The dashed line is a theoretical fit with a model using the two units of optical quantum noise and the finite mechanical evolution. The solid line is the inferred conditional mechanical width immediately after the preparation pulse. The vertical line indicates the pulse strength used for states (a-c). 
}
\label{Fig:Results}
\end{figure}

\end{widetext}

\section*{Mechanical motional state reconstruction and state preparation via measurement}

The mechanical resonator used for this experiment is a micro-mirror cantilever constructed from an epitaxial Al$_{\textrm{x}}$Ga$_{1-\textrm{x}}$As crystalline multilayer, see Fig.~1(b). The use of such a monocrystalline material structure allows for a significant reduction of the mechanical damping of the resonator when compared with dielectric reflectors~\cite{Harry2002} and simultaneously provides high optical reflectivity. The crystalline material used here is nominally identical in composition and individual layer thickness to the structures in Ref.~\cite{Cole2010} and is designed for maximum reflectivity at our optical wavelength of 1064 nm. The multilayer mirror comprises 40.5 layer pairs in order to minimize transmission losses. The cantilever was etched from a 6.88~$\mu$m thick multilayer and is 1.45~mm in length with a cantilever arm 5~$\mu$m in width with a circular head 100~$\mu$m in diameter where the optical signal beam is focussed~\cite{Cole2012}. Note that the resonator is etched directly from the multilayer mirror material and is therefore equally reflective at all points along the structure with an (intensity) reflectivity of 99.982~\%. This cantilever has a fundamental out-of-plane vibrational mode with frequency $\omegam/2\pi = 984.3$~Hz, effective mass $\meff = 260$~ng \cite{Supplementary}, ground state extension $\xzp = 5.7 \times 10^{-15}$~m, and a mechanical quality of $Q = 3.1 \times 10^4$ in vacuum ($10^{-5}$~mbar) and at room temperature measured via mechanical ringdown.

Our optical setup (Fig.~1(a)) was constructed from optical-fiber-based components that provided good phase stability and excellent spatial mode matching. Indeed, when the optical powers in the two arms of the interferometer are balanced we observed an interference visibility exceeding 99.9~\%. We use a continuous laser source and generate optical pulses of duration 1~$\mu$s (excluding the pulse edges) with a fiber-based intensity modulator. The mean photon number in a signal pulse was up to $10^{7}$ and in order to provide a homodyne signal well above the electronic noise we use a large LO to signal ratio with up to $10^{10}$ photons per LO pulse. (These photon numbers were determined via optical power measurement during continuous wave operation.) The signal pulses are directed onto the cantilever head using an anti-reflection coated fiber focuser and are then retro-reflected. To calibrate the proportionality between the measurement outcomes and the mechanical position we reflect the signal beam from a rigid mirror adjacent to the mechanical resonator and scan the mirror position using a calibrated piezoelectric actuator recording both the piezo scan positions and measurement outcomes~\cite{Supplementary}. For our mechanical resonator ground state extension, this photon number per pulse yields a measurement strength $\chi$ of order $10^{-4}$ and a momentum transfer $\Omega$ of order unity. The radiation pressure backaction from the reflection of the pulse is smaller than the mechanical thermal noise and is not observed, however, as will be detailed in the following, this measurement strength has a strong effect on the mechanical thermal noise.

After a pulse measurement is performed to sample a mechanical marginal, the mechanical state is reinitialized by first allowing it to return to equilibrium with the environment and then the mechanical state is re-prepared. This process is repeated many times to accumulate sufficient data to characterize the statistical properties of the mechanical motion. The marginal distributions were then obtained by constructing a histogram from the many measurement outcomes recorded for each mechanical phase-space angle $\theta$. As the states studied here are symmetric about the $\XM$ and $\PM$ axes we measure a set of many marginals with angles between $\theta = 0$ and $\theta = \pi/2$ to fully characterize the state of motion. The phase-space probability distribution $W(\XM, \PM)$ is then obtained by using the inverse Radon transformation on the set of marginals.

The measurement results we obtained for motional state preparation and reconstruction are summarized in Fig.~\ref{Fig:Results}. In Fig.~\ref{Fig:Results}(a) a reconstruction of an initial thermal state that is driven by white noise across the mechanical resonance up to a mode temperature of 1100~K that has width $\sigma_{x} = 1.2$~nm is shown. This temperature was obtained using the equipartition theorem $k_B \Teff = \meff \, \omegam^2 \sigma_{x}^2$, where the mechanical position variance $\sigma_{x}^2$ was obtained from the calibrated measurement outcome distribution after subtracting the optical noise contribution.
A single pulsed measurement made on this initial thermal state generates a motional state that has a reduced position uncertainty (Fig.~\ref{Fig:Results}(b)). The observed momentum distribution of this state, however, is unchanged as the back-action to the mechanical momentum made by the reflection of the optical pulse is much smaller than the mechanical thermal noise. Each pulsed measurement generates a mechanical state with a random but known mean due to the random measurement outcome, see Eq.~(\ref{eq:mechInOut}). By making the transformation $\PLRead \rightarrow \PLRead - \PLOne \cos\theta$, where the superscripts $(r)$ and $(p_1)$ indicate read-out and preparation, respectively, this random mean is subtracted and the distribution of the mechanical state can be characterized. We would like to emphasize here that no `post-selection' is performed and all measurement outcomes are used in this process. Furthermore, our experimental pulsed technique demonstrates the back-action-evading feature of measurement repeatability - a subsequent measurement is not affected by a prior measurement~\cite{ref:QND,Braginsky1980}. Specifically, in our case the measurement results of the read-out pulse made a short time after the preparation pulse are the same as the preparation pulse to within the optical quantum noise. The plots for Figs. \ref{Fig:Results}(a,b) were generated from the same data set where the statistics of the preparation pulse alone characterizes the unconditional initial thermal state and the read-out pulse characterizes the conditional mechanical state. A 1100~K thermal state (which has an RMS amplitude less than two larger than a thermal state at 300~K) was used to increase the mechanical contribution to the optical phase noise over the relevant $\sim$ DC to MHz bandwidth for our pulses to improve the signal-to-noise ratio for mechanical conditional state preparation.

In Fig.~\ref{Fig:Results}(c) the reconstruction of a mechanical state of motion prepared via two pulsed measurements separated by one quarter of a mechanical period is shown. The width of the mechanical phase-space distribution has been significantly reduced in both the position and momentum quadratures compared to the initial thermal state (Fig.~\ref{Fig:Results}(a)) and hence the effective mode temperature has decreased. This method of cooling is rapid as it takes place well within a single mechanical period and is, to the best of our knowledge, yet to be experimentally reported elsewhere. For this pulse sequence the read-out pulse outcome is transformed using $\PLRead \rightarrow \PLRead - \PLTwo \cos\theta + \PLOne \sin\theta$, where $\theta$ is the angle of mechanical evolution made between the second preparation pulse and the read-out pulse. Ideally, for this mechanical state, the width of the mechanical marginals should be constant for all $\theta$, however, in our experiment the phase correlation between the pulses reduces with increasing pulse separation as low frequency noise, due to imperfect phase locking, enters the signal. This results in a broadening of the conditional mechanical marginals as $\theta$ increases. The effective temperature $\Teff = \meff \, \omegam^2 \sigma^{(\theta = 0)}_x \sigma^{(\theta = \pi/2)}_x / k_B$ observed for this state is 16~K, which depends on the product of the standard deviations of the position and momentum quadratures. Were the pulses to remain correlated to within the quantum noise, the effective temperature that could be reached for this measurement strength, taking the effects of mechanical rethermalization into account is 4.4~K~\cite{VannerPNAS2011}. We would like to highlight here that rethermalization contributes to only less than 1~\% of this value. To summarize the observed effects of single and two-pulse mechanical state preparation discussed above, Fig.~\ref{Fig:Results}(e) provides a plot of the measured mechanical widths with $\theta$ for the initial thermal state and the two mechanical conditional states. In this plot the mechanical widths were determined from the calibrated pulse outcome distributions after subtracting the optical noise contributions that were measured independently. The data for both of the mechanical conditional states was taken with the same signal pulse powers and for each phase space angle 300 pulses were recorded to construct the histograms. 


As an example of a non-Gaussian state of motion we have reconstructed a driven thermal state (Fig.~\ref{Fig:Results}(d)) that was generated by applying a sinusoidal drive on resonance with the mechanical eigenfrequency. Note that the two peaks in the mechanical marginals are narrower than the broad thermal state in Fig.~\ref{Fig:Results}(a) as this state was prepared at room temperature without the white noise drive. Even though this state of motion and the thermal state are rotationally invariant in phase-space many marginals are measured for their reconstruction. On the other hand, the conditional mechanical states of motion are not rotationally invariant in phase-space as the time of the preparation pulse(s) sets the time for $\theta = 0$. Note that this pulse-based tomography scheme does not measure the angle $\theta = 0$ as the read-out pulse is temporally separated from the preparation pulse(s). The lack of this marginal angle causes the rippling near $\XM = 0$ in the reconstructed phase-space distributions. By employing shorter pulses and measuring the marginals at smaller angles this rippling can be reduced.

To demonstrate the scaling of our measurement strength in Fig.~\ref{Fig:Results}(f) the conditional mechanical width observed by a read-out pulse made after $5^\circ$ of mechanical free evolution is plotted with increasing pulse amplitude. For this pulse separation the two pulses are well correlated and the width of the conditional mechanical state is limited by the optical quantum noise in the measurement (see the methods section for more details). As the signal pulse strength is increased the standard deviation of the conditional mechanical position distribution decreases with $N^{-1/2}$, which is a result of the optical number-phase uncertainty relation. The dashed line in the plot is a theoretical prediction including the two units of optical shot noise, one each for the preparation and read-out pulses, and the small contribution from the mechanical evolution between the two pulses. The relative amplitudes for the data points were measured precisely and scaled by a free fitting parameter into units of square-root photon number, where the photon number per pulse obtained is consistent with measurements of the optical power made during continuous wave operation. For the largest optical pulse strength used the statistics of the read-out pulse  demonstrate a conditional mechanical width (after the preparation pulse) of $\sigma_x = 19$~pm corresponding to a measurement strength of $\chi = 2.1 \times 10^{-4}$.

\section*{Discussion}

The techniques developed in this work provide the ability to experimentally perform quantum optomechanics in the time domain. This offers significant potential for optomechanics-based quantum information and quantum metrology applications by providing the framework for quantum state preparation of a mechanical resonator via quantum measurement~\cite{Wiseman2010}. One may then also envision combining such measurement based state preparation with feedback to implement full quantum control~\cite{Sayrin2011}. One exciting example of mechanical dynamics that can be probed by pulsed optomechanics has been recently theoretically discussed by Buchmann \emph{et al.}~\cite{Buchmann2012}, where pulsed measurements, as now realized in this work, are considered for the observation of quantum tunneling of a mechanical oscillator in a double-well potential. Another example for quantum state preparation is that, even though the optomechanical interaction used here is linear with the mechanical position, by exploiting the optical non-linearity, $X_M^2$ measurements with a strength significantly larger than that attainable with dispersive optomechanics can be performed~\cite{VannerPRX2011}. An $X_M^2$ measurement can be used to conditionally prepare highly non-Guassian mechanical superposition states and experimentally characterizing the decoherence of such states is important to determine the feasibility of using mechanical elements for coherent quantum applications and can also be used to empirically test collapse models \cite{Bose1999, Marshall2003, Kleckner2008, RomeroIsart2011}. The pulsed measurements performed here may also be utilized for a QND-measurement-based light-mechanics quantum interface~\cite{Marek2010}. Furthermore, a sequence of four pulsed optomechanical interactions can be used to generate non-classical mechanical states of motion via an optomechanical geometric phase~\cite{Khosla2012} and can even be used to experimentally explore potential quantum-gravitational phenomena~\cite{Pikovski2012}.

For this experiment, to prepare a quantum squeezed state of mechanical motion the measurement strength needs to be increased to $\chi > 1$. An effective route to meet this requirement would be to employ an optical cavity to enhance the optomechanical interaction. Using the experimental parameters achieved in this work, a cavity finesse of $10^4$ is sufficient. As such a cavity simultaneously requires a high finesse, as well as a large bandwidth to accommodate a short optical pulse, this is best achieved with an optomechanical microcavity~\cite{VannerPNAS2011}. Such improvements to the measurement sensitivity will not only enable Wigner reconstruction with significant negativity but, owing to this pulsed protocol's resilience against mechanical thermal noise, may also allow the generation of non-classical mechanical states in the regime of room temperature quantum optomechanics.

\section*{Methods}

\begin{figure}[!ht]
\includegraphics[width=0.9\hsize]{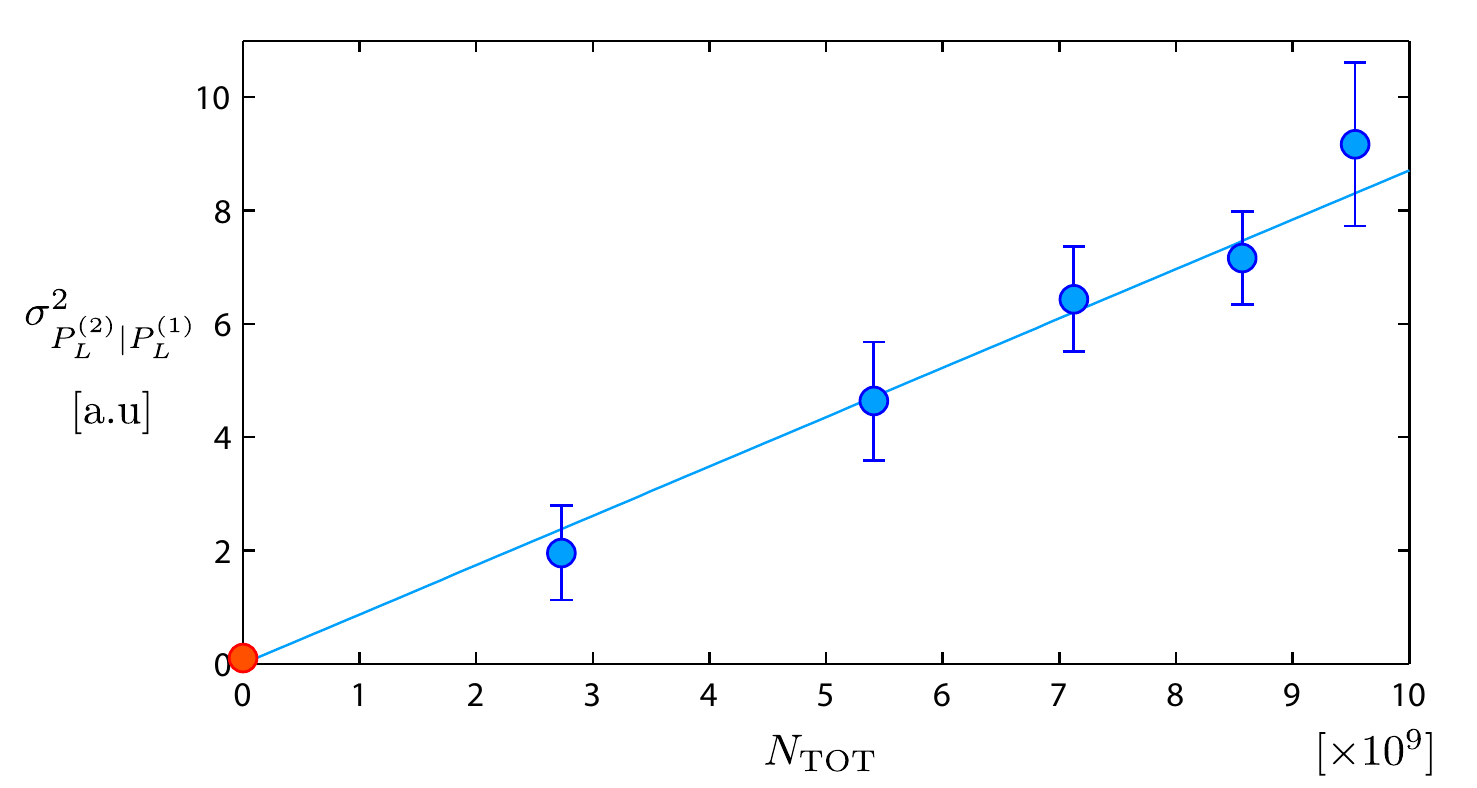}
\caption{
The measured optical phase quadrature conditional variance $\sigma^2_{P_L^{(2)}|P_L^{(1)}}$ plotted with the total photon number (signal plus local oscillator) per pulse $N_{\textrm{TOT}}$. The linear dependence observed demonstrates that the measurement scheme is optical quantum noise limited up to $N_{\textrm{TOT}} \simeq 10^{10}$. Were classical phase noise to be the dominant noise source a quadratic dependence would be observed.
}
\label{Fig:QuantNoise}
\end{figure}

To verify that the measurement scheme used here is optical quantum noise limited we measured the phase quadrature conditional variance of a pair of optical pulses with increasing total photon number, i.e. the sum of the signal and LO photons per pulse, while keeping the signal to LO ratio fixed, see Fig.~\ref{Fig:QuantNoise}. As with our calibration procedure, the signal beam is focussed onto a rigid mirror adjacent to the mechanical oscillator to prevent coupling to the mechanical motion. The pulse separation used for this measurement was $14.1~\mu$s, which would correspond to $5^\circ$ of mechanical free evolution, and is the same as that used for the dataset shown in Fig.~\ref{Fig:Results}(f). With this pulse separation the conditioning is essentially the second pulse outcome minus the first pulse outcome. The quantum noise components of these two temporal modes are uncorrelated, however, the lower frequency classical noise components vary slowly between the two pulses and are thus suppressed by the conditioning. Quantum mechanics predicts a linear dependence for the variance with total photon number, whereas, were classical phase noise to be the dominant contribution, a quadratic dependence with the total photon number per pulse would be observed~\cite{Bachor2004}. During this measurement we were limited to a total photon number of $10^{10}$ as the phase lock performance dramatically reduced beyond this point. Were we able to measure beyond this optical power the classical phase noise would eventually become the dominant noise and the conditional mechanical variance that can be achieved would saturate.

The data points for Fig.~\ref{Fig:QuantNoise} were obtained from Gaussian fits to histograms of the conditional outcomes. The error bars indicate a one standard deviation uncertainty as determined from the fit. The observed conditional variance shows a linear dependence with the total photon number with a `goodness of fit' parameter $R^2 = 0.97$, taking the error bars into account. This demonstrates that, up to a total photon number of order $10^{10}$, the conditional variance is quantum noise limited.

Also included in Fig.~\ref{Fig:QuantNoise} is the measured electronic noise, i.e. the conditional variance observed using no light. This contribution is 19.5~dB smaller than the observed optical quantum noise at the data point with the highest optical intensity ($N_{\textrm{TOT}} = 9.5\times10^{9}$).

\phantom{blank line}

\section*{Acknowledgments}

We thank K.~Hammerer, S.~G.~Hofer, M.~S.~Kim, G.~J.~Milburn, I.~Pikovski, R.~Riedinger, and J.~Schm\"{o}le for useful discussion. M.R.V. is a
member of the FWF Doctoral Programme CoQuS (W 1210) and is a recipient of a DOC fellowship of the Austrian Academy of Sciences. Microfabrication was carried out at the Zentrum f\"{u}r Mikro- und Nanostrukturen (ZMNS) of the Technische Universit\"{a}t Wien and the epitaxial multilayer was grown by Markus Weyers' group at the The Ferdinand-Braun-Institut, Leibniz-Institut f\"{u}r H\"{o}chstfrequenztechnik (FBH), Berlin (Germany). We thank support provided by the European Comission (Q-ESSENCE, CQOM), the European Research Council (ERC QOM), and the Austrian Science Fund (FWF) (START, SFB FOQUS).


\newpage
\phantom{a}
\newpage

\appendix
\section*{Supplementary Information - Experimental Pulsed Quantum Optomechanics}

\subsection*{Effective mass measurement}

\begin{figure}[!ht]
\includegraphics[width=0.9\hsize]{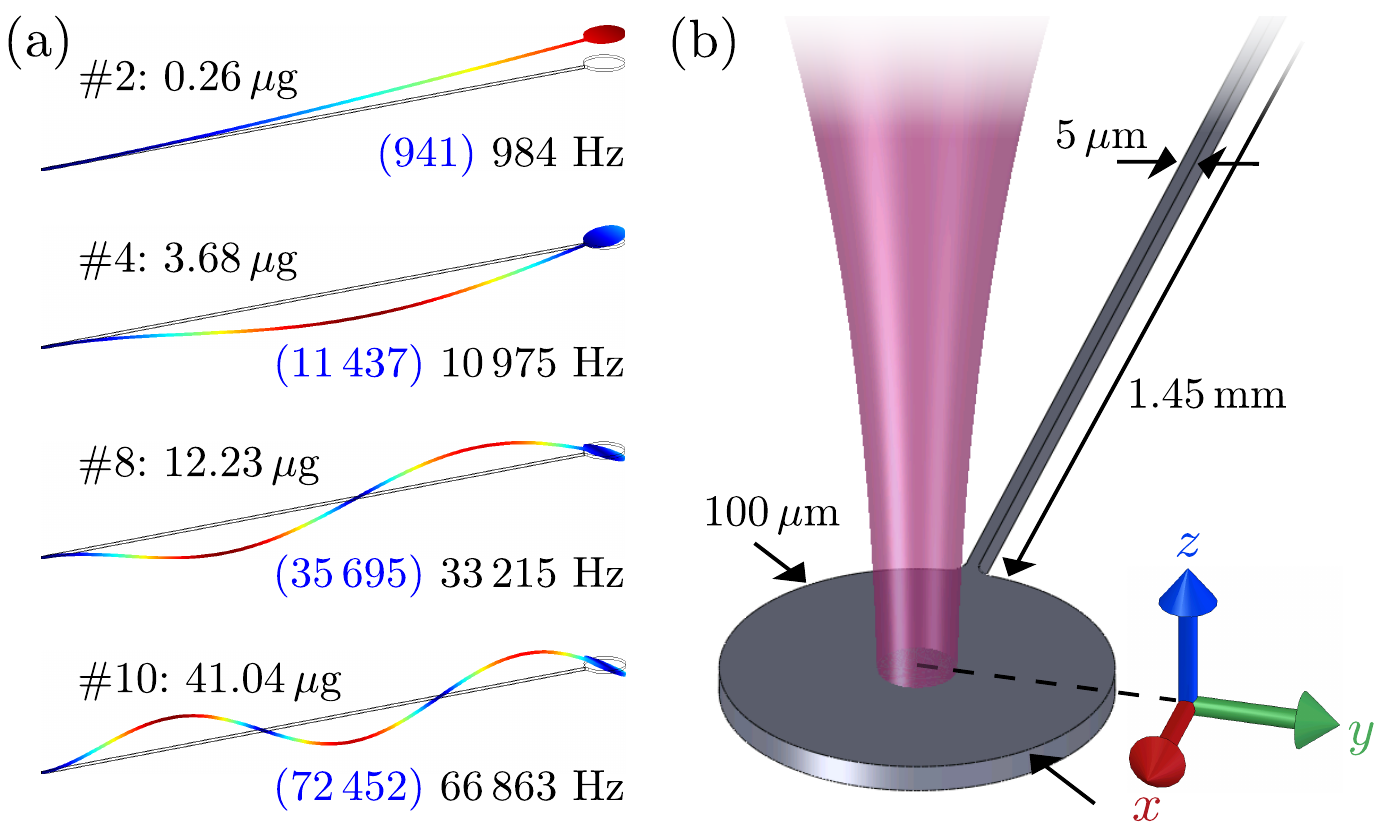}
\caption{
(a) Finite element simulations of the vibrational modes ($\#2,4,8,10$) that have the four lowest optically probed effective masses. The displacements shown are exaggerated (blue to red indicates increasing displacement) and the outline indicates the cantilever rest position. The optically probed effective mass is given next to each mode number and underneath are the simulated mechanical frequencies in brackets and the measured frequencies. (b) Schematic of the cantilever and the focussed signal beam. The coordinate axis used when determining the optically probed effective mass is in the center of the cantilever head on the mirror surface.
}
\label{Fig:Mechanics}
\end{figure}

The optically probed effective mass of a mechanical vibrational mode depends upon (i) the geometry and material properties of the mechanical structure and (ii) the intensity profile of the incident optical beam. The mass associated with the mechanical displacement mode shape, i.e. the modal mass, is in general less than the total mass of the structure, however, the optically probed effective mass can have a strong dependence on the position and profile of the optical beam. We estimate the optically probed effective mass of the cantilever in our experiment using a combination of measurements and finite element analysis (FEA). Using the established values for the relevant elastic constants averaged over the crystalline multilayer ($C_{11} = 119.6$, $C_{12} = 55.5$, $C_{44} = 59.1$ GPa) and the average material density 4476 kg/m$^3$, the lateral geometry of the FEA-simulated resonator is adjusted until minimal error is found between the measured and simulated eigenfrequencies for the first four out-of-plane mechanical modes, see Fig.\ref{Fig:Mechanics}(a). (Note that the lowest frequency vibrational mode for our cantilever is an in-plane mode as the cantilever used is slightly thicker than wide.) A mean discrepancy between the measured and simulated frequencies of 6.1\% was obtained by reducing the feature linewidth by 0.875~$\mu$m with respect to the lithographic mask. Note that the thickness of the free-standing mirror material was not used as a fitting parameter as it was accurately determined from the reflectance spectrum of the mirror~\cite{Cole2011} and found to be 6.88~$\mu$m. Once the geometry is determined, the effective mass is calculated via the volume integral \cite{Pinard1999}
\begin{equation}
\meff = \frac{\rho \int\!\!\int\!\!\int\!\textrm{d}x\,\textrm{d}y\,\textrm{d}z \left(u^2 + v^2 + w^2 \right)}{\mathcal{D}^2}.
\end{equation}
Here, $\rho$ is the material density; $u$, $v$, and $w$ are the displacements of the body along the $x$, $y$, and $z$ directions, respectively, and the optically-probed displacement $\mathcal{D}$ is the overlap between the mechanical deflection and the optical Gaussian intensity profile, i.e.
\begin{equation}
\mathcal{D} = \frac{1}{2\pi r_0^2}\!\int\!\!\!\int\!\textrm{d}x\,\textrm{d}y \,\, w(x,y,z\,{=}\,0) \exp\!\left[ - \frac{x^2 + y^2}{2r_0^2} \right],
\end{equation}
where $r_0$ is the standard deviation of the Gaussian optical intensity profile and the coordinate axis used for $x$, $y$ and $z$ has its origin in the center of the cantilever head, see Fig.~\ref{Fig:Mechanics}(b). The anti-reflection coated fiber focuser used in our experiment provides an optical beam diameter ($4r_0$) of $10.6~\mu$m, which is much smaller than the nominal cantilever head diameter of 100~$\mu$m (as fabricated diameter of 98.25~$\mu$m). Thus, for the fundamental out-of-plane mode, there is only a weak optical beam position and width dependence on the effective mass. (In this case, the effective mass is approximately equal to the intrinsic modal mass.) We have determined that the fundamental out-of-plane mechanical mode utilized in our experiment, which oscillates at 984.3~Hz, has an effective mass of 260~ng and thus a spring constant of 0.01~N/m. For the higher order modes of the structure, however, lateral displacement of the beam leads to a rapid change in the effective mass. In order to minimize the contribution from these higher mechanical modes it is necessary to carefully position the optical beam. Assuming careful alignment, the geometry of our mechanical structure is such that the contributions from higher order modes are further suppressed as the effective mass rapidly increases with mode number since the majority of the deflection is due to the support beam for the fundamental mode rather than in the head for higher harmonics. Indeed, the unconditional RMS amplitudes of modes $\#4$, $\#8$, and $\#10$ are 2.4\%, 0.4\%, and 0.1\% that of mode $\#2$, respectively.

\subsection*{Calibration Procedure}

We have used a two-step calibration procedure to determine the proportionality between the pulsed homodyne measurement outcomes and the mechanical displacement. During this procedure the signal beam is focused onto the chip edge, i.e. a rigid unpatterned part of mirror material adjacent to the mechanical resonator, to prevent mechanical motion contributing to the signal. First we calibrate the displacement of a piezoelectric actuator, which our fabricated structure containing the mechanical oscillator is placed upon, in response to a known drive voltage. We then drive the piezo and record the pulse measurement outcomes during the controlled actuation in order to calibrate the pulsed interferometer. Each step is detailed below in the next two subsections, respectively.

\begin{figure}[!ht]
\includegraphics[width=0.9\hsize]{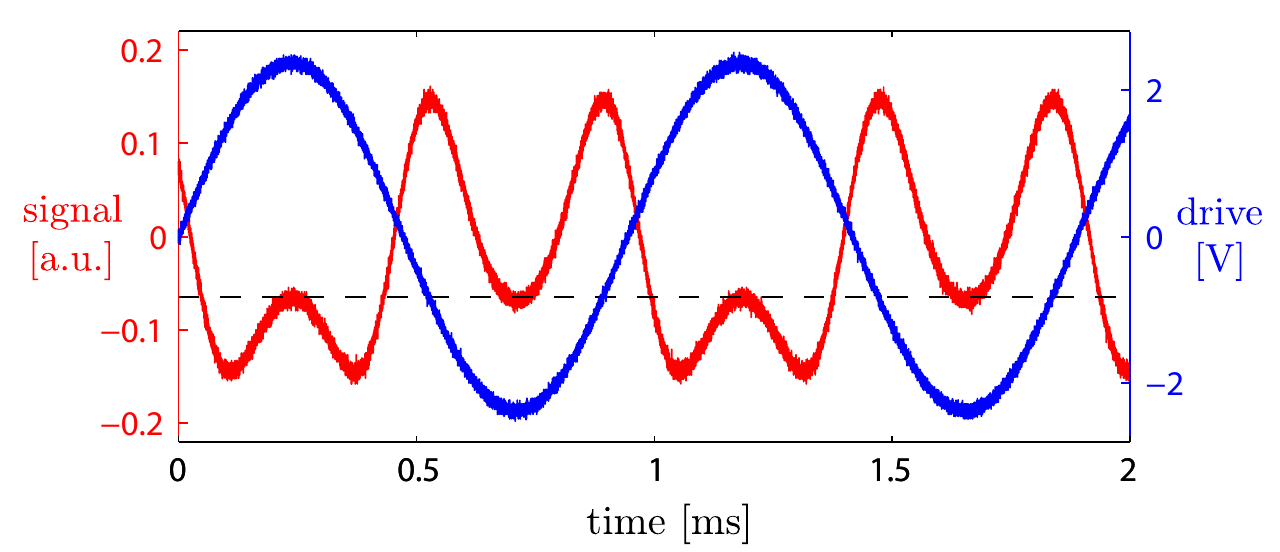}
\caption{
Plot of the homodyne output (red, left axis) during piezo drive (blue, right axis) to calibrate the piezo response. The sinusoidal drive generates a peak-to-peak piezo scan of one half of an optical wavelength that was determined by adjusting the drive amplitude until the turning points in the homodyne output coincide (dashed line).
}
\label{Fig:Calibration}
\end{figure}

\subsubsection*{Piezo Calibration}

To calibrate the piezoelectric actuator we applied a sinusoidal drive voltage and used a continuous signal beam to monitor the piezo motion. The frequency of the drive was chosen such that the piezo mechanical response was either in or out of phase with the drive voltage. (Experimentally, care was needed to find a suitable drive frequency as the piezo does not have a flat spectral response.) During this procedure, the phase between the signal and LO beams does not require locking and the piezo drive was at a higher frequency than the phase noise components in the interferometer. We then adjusted the drive amplitude such that the peak-to-peak piezo motion was one half of the optical wavelength. This can be done precisely as the difference current output of the interferometer has separate turning points occurring at the same level for this modulation depth and is then proportional to $\cos[\phi_0 + \pi \sin \omega t]$, see Fig.~\ref{Fig:Calibration}, here $\phi_0$ is the (unlocked) slowly varying phase in the interferometer and $\omega$ is the piezo drive angular frequency. As $\phi_0$ slowly changes this merely shifts the level of the turning points. In our experiment we used a drive frequency of 1.06~kHz and exploited a resonance of the piezo to achieve a peak-to-peak scan of 532~nm using 4.6~Vpp.

\subsubsection*{Pulse Calibration}
Using the same piezo drive frequency as above, and using the piezo actuator calibration value (meters per Volt) obtained, the actuator was scanned with a reduced amplitude so that the optical phase shifts are small. (It was verified that the piezo responds linearly with the applied Voltage over our range of interest.) Then, during the piezo scan, pulsed position measurements are performed and both the voltage applied to the piezo at the time of the measurement and the pulsed measurement outcomes are recorded. The proportionality between these recorded values is used to obtain the outcome per meter calibration. This calibration value is optical amplitude dependent and had to be measured for several optical amplitudes for the measurement shown in Fig.~\ref{Fig:Results}(f) of the main text.

\end{document}